\documentclass[twocolumn,prl,aps]{revtex4}
\usepackage{graphicx}

\begin{document}

\title{Unified description of \emph{quantum nonlocal} and \emph{relativistic local} correlations:\\
Both assume ``free will" and happen without connection in space-time.}

\author{Antoine Suarez}
\address{Center for Quantum Philosophy \\ Ackermannstrasse 25, 8044 Zurich, Switzerland\\
suarez@leman.ch, www.quantumphil.org}

\date{October 3, 2015}

\begin{abstract}

It is argued that quantum and relativistic correlations can be described in a unified way, in that both assume ``free will" as an axiom, and happen without any continuous connection in space-time. This description may contribute to a coherent definition of ``space-time quantization" and highlights the importance of solving the ``measurement problem".

\ \\
\textbf{Keywords:} Quantum correlations, relativity, wavefunction collapse, decision at detection, free will, trajectories, quantization of space-time, measurement problem.

\end{abstract}

\pacs{03.65.Ta, 03.65.Ud, 03.30.+p}

\maketitle

\ \\
\textbf{Introduction}.\textemdash Standard quantum mechanics assumes that the decision of the outcome happens at the moment of detection (``wavefunction collapse''). At the 5$^{th}$ Solvay conference (1927) Einstein objected to this assumption by means of a \emph{single-particle} gedanken-experiment. The quantum collapse, he argued, implies a \emph{nonlocal} coordination of the detectors, which cannot be explained by influences propagating with velocity $v\leq c$; this involves ``an entirely peculiar mechanism of action at a distance, which [...] implies to my mind a contradiction with the postulate of relativity." \cite{bv}

Astonishingly Einstein's gedanken-experiment in 1927 has been first realized in 2012.\cite{Guerreiro12} This experiment demonstrates nonlocally coordinated detector's behavior, and also highlights something Einstein did not mention: Nonlocality is necessary to preserve such a fundamental principle as energy conservation.\cite{Guerreiro12} Regarding Einstein's claim that there is ``a contradiction" between quantum nonlocality and relativity, it has been argued that the ``contradiction" exists only in ``Einstein's mind", that is, in \emph{his} interpretation of relativity: In fact, quantum physics and relativity are confirmed by one and the same experiment, the single-photon space-like Michelson-Morley experiment; this experiment demonstrates that quantum physics and relativity imply each other.\cite{as14}

\begin{figure}[t]
\includegraphics[trim = 10mm 182mm 10mm -6mm, clip, width=0.99\columnwidth]{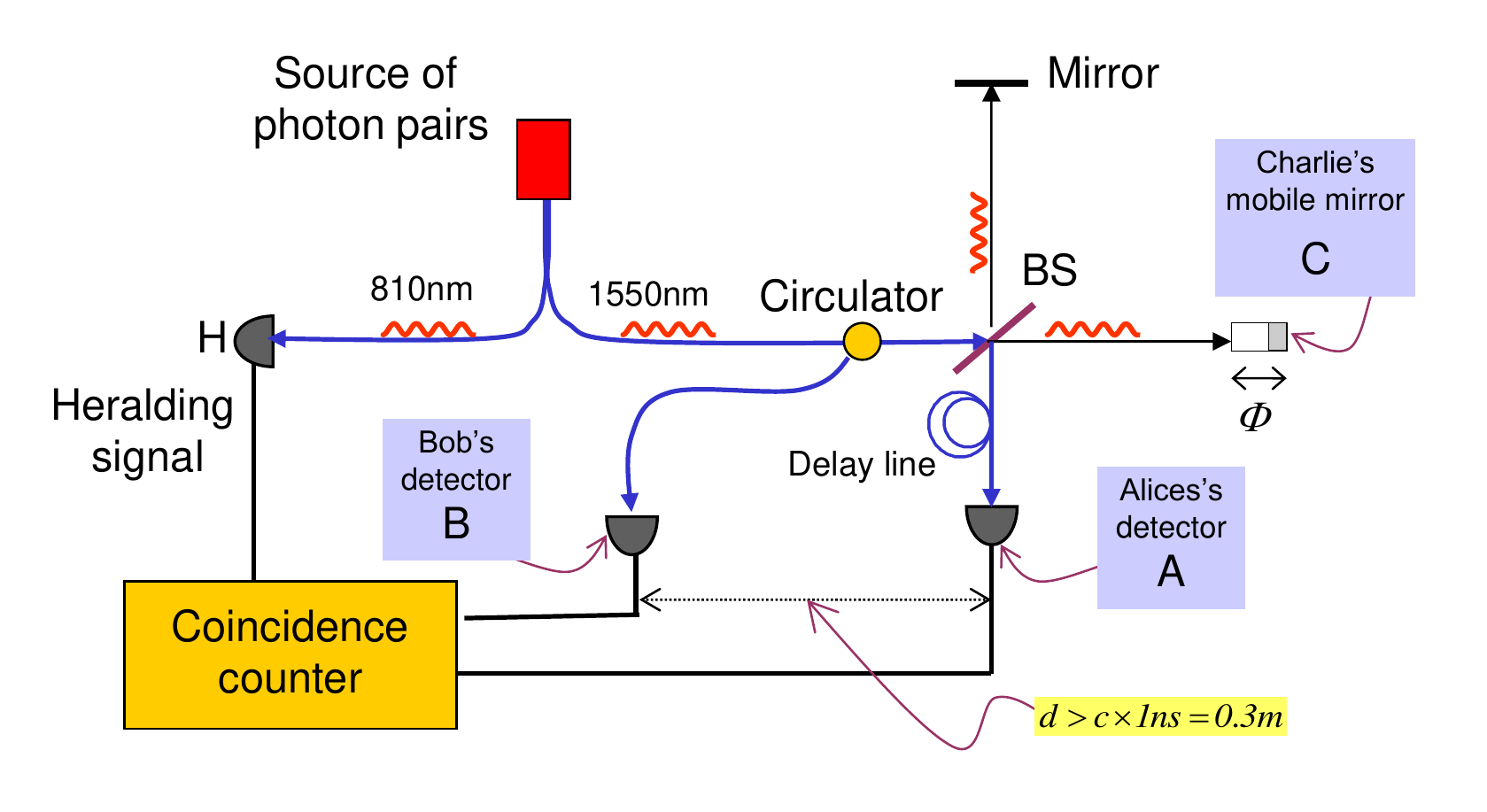}
\caption{\textbf{Single-photon space-like Michelson-Morley experiment:}
It tests with single photons that the velocity of light cannot be measured in an absolute way (to this aim Charlie's arm is first oriented in the direction of the Earth's motion, and subsequently the interferometer is rotated 90$^\circ$) \cite{as14}; the very same experiment tests that the decisions at Alice's detector A and Bob's one B (i.e.: ``A fires" and ``B doesn't fire", or viceversa) are strictly correlated and space-like separated (to this aim the detectors A and B are conveniently separated from each other) \cite{as14}: The quantum correlation between the decisions at detectors A and B is \emph{nonlocal} and happens without connection in space-time; the correlation between the settings of Charlie's mirror and the counting rates of Alice's (or Bob's) detector is relativistic \emph{local}, nonetheless it happens without well-defined trajectory or continuous connection in space-time as well.}
\label{f1}
\end{figure}

In this letter I take a further step and argue that quantum and relativistic correlations can be described in a unified way: On the one hand ``free will" is an axiom of both, quantum physics and relativity; on the other hand quantum and relativistic correlations happen without any continuous connection in space-time.

\ \\
\textbf{The single-photon space-like Michelson-Morley experiment}.\textemdash I first summarize the single-photon space-like Michelson-Morley experiment as presented in \cite{as14} and sketched in Figure \ref{f1}.

A source produces pairs of photons and one of them is used for heralding, i.e. to signaling the presence of a photon in the interferometer and opening the counting gate, as indicated in \cite{Guerreiro12}. The other photon enters the interferometer through the beam-splitter BS and, after reflection in the mirrors, leaves through BS again and gets detected. Each interferometer's arm is supposed to have equal length $L$.

Such interference experiments can be considered the entry into the quantum world. With sufficiently weak intensity of light, only one of the two detectors clicks: either A or B (\emph{photoelectric effect}). Nevertheless, for calculating the counting rates of each detector one must take into account information about the two paths leading from the laser source to the detector (\emph{interference effect}). If $a \;\in\{+1,-1\}$ labels the detection value, according to whether detector A or detector B clicks, the probability of getting $a$ is given by:

\begin{footnotesize}
\begin{eqnarray}
P(a)=\frac{1}{2}(1+ a \cos \mathit\Phi)
\label{Pa}
\end{eqnarray}
\end{footnotesize}

\noindent where $\mathit\Phi=\omega\tau$ is the phase parameter, $\omega$ the angular frequency, and $\tau$ the optical path difference i.e., the difference between the times light take to travel each path of the interferometer. If one assumes the same light speed in the two paths of the interferometer, then $\tau=\frac{l-s}{c}$. The phase $\mathit\Phi$ can be changed by means of a mobil mirror C.

As said, according to standard quantum mechanics which detector clicks (the outcome) is decided by a choice on the part of nature when the information about the two paths reaches the detectors. And this implies a coordinated behavior on the part of A and B, no matter how far away from each other these detectors are.

On the other hand the experiment in Figure \ref{f1} can be used to perform a new version of the Michelson-Morley experiment \cite{MM} that uses single-photons and two detectors A and B (instead of only one).\cite{as14}

In this context the assumption of ``a timeless quantum collapse" means that it is impossible to define the velocity of light in an absolute way with relation to a preferred referential frame. And this means that quantum physics not only does not contradict relativity, but rather implies it.\cite{as14} On the other hand, if the detectors A and B are conveniently set /as indicated in Figure \ref{f1}) \cite{Guerreiro12}, the confirmation of relativity by a negative Michelson-Morley result would show that the decisions at A and B (i.e.: ``A fires" and ``B doesn't fire", or viceversa) are space-like separated, and thereby demonstrate the nonlocal nature of the quantum collapse: The correlated behavior exhibited by A and B cannot be explained by influences propagating with velocity $v\leq c$; the quantum collapse implies relativity, and the relativistic structure of the space-time implies that the quantum collapse is nonlocal (i.e., comes from outside space-time) and cannot be used by the experimenter for communication faster than light.\cite{as14}

Although the experiment in \cite{as14} has not yet been done, reviewers have suggested that he can be considered done on the basis of the original Michelson-Morley experiment \cite{MM} and the single-photon space-like antibunching one \cite{Guerreiro12}. Anyway, a refreshed negative Michelson-Morley result would show that quantum physics and relativity imply each other: you cannot have one without the other.

\ \\
\textbf{Free will is an axiom of both, quantum physics and relativity}.\textemdash Regarding quantum physics: Suppose the photon behaves like a classical particle and cross the interferometer of Figure \ref{f1} by one of the two arms, following a well-defined trajectory in space-time; suppose that Charlie is contrived by nature to move his mirror only when the photon takes this path; then the photon can always carry the information about the optical path difference, and one could explain interference without any need of the characteristic quantum principles. Hence ``the \emph{free} will" of the experimenter is an axiom of quantum physics and ``comes first in the logical order" \cite{ng}.

Regarding relativity: Suppose the source emits two sorts of photons, ``fast photons", which take the arm set in the direction of the movement of the Earth, and ``slow photons", which take the other arm. Such an hypothesis would explain the negative Michelson-Morley result without need of the relativity principle. This objection can be countered by assuming that Charlie may \emph{freely} change the length of his path after the photon has entered the interferometer, and thereby thwarts any possible pre-selection of photons at the source. Accordingly ``free will" is in fact also an axiom of relativity.

\ \\
\textbf{Quantum and relativistic correlations happen without any continuous connection in space-time}.\textemdash The standard view of decision at detection has to main implications:

1) Photons do not cross the interferometer taking a well-defined path or trajectory through space–time.

2) Nonlocal coordination of the firings at the detectors A and B; this quantum correlation cannot be explained by any mechanism in space-time, that is by a continuous flow of particles (signals) traveling between A and B.

Suppose Charlie changes the settings of the mobile mirror C in the setup of Figure \ref{f1}, and consequently the firing rates of the detectors A and B. Thereby he produces a correlation between the mirror's settings and the behavior of the detectors, and can send a message to Alice and Bob. Such a procedure is a paramount example of relativistic local causality that can be used for communication through signaling (e.g.: phoning).

Nonetheless, since the photons do not follow any trajectory in the interferometer (implication 1 above), we have to conclude that the relativistic local correlation between the C settings and the A counting rates happens without any continuous connection (continuous flow of particles) between C and A.

Similarly one can state that  between the activation of the laser source and the detector's firing there is a relativistic local correlation, which happens without photons traveling through well-defined trajectories in space-time.

Suppose one enlarges one of the arms of the interferometer beyond the coherence length of the laser source. In this case one could describe things by stating that the photon travels either the long or the short pat. But does this mean that trajectories suddenly appear? In my view, the coherent explanation, even in this case, is that the photons do not cross the interferometer by following any well-defined trajectory: ``Trajectories" (and therefore ``particles") are only conceptual constructs useful for calculus. Quantum physics is all about correlations between observed events.

This explanation can be generalized to any correlations between visible events, as for instance activation of a source and detection, even without any interferometer in between. Activating the source is certainly a necessary condition in order the detector counts, but this does not imply any continuous trajectory in space-time between source and detector.

If one assumes a continuous flow of particles going on between a source and a detector, one cannot escape empty waves, and many-worlds at the end, that is, the assumption of entities or regions that exist within space-time and nonetheless are inaccessible (invisible).\cite{Guerreiro12} In my view such an assumption is not coherent. If some entity is supposed to be invisible, it cannot be within space-time, since the ``real physical space-time" is nothing other than the ensemble of correlated observations.

By contrast, one can coherently assume that the correlations between observed events (no matter whether they are quantum nonlocal or relativistic local ones) originate always from outside space-time: There are correlated visible things, which define the ``real physical space-time", and invisible influences producing the correlations from outside space-time!

\ \\
\textbf{Space-time quantization and the ``measurement problem"}.\textemdash This unified description of quantum nonlocal and relativistic local correlations can contribute to a coherent definition of space-time quantization and thereby overcome the inconsistencies of general relativity:

The ``quantum of space-time" is not necessarily defined by the Planck scale but by the ``minimal distance" two observed events can have. And to define this ``minimal distance" it is necessary to define sharply what an observation is and when does it happen, that is, to solve ``the measurement problem".

This problem is certainly at the heart of quantum physics. But it does not mean at all that a human experimenter has to be watching a detector in order it decides to fire or not, but rather this: The process by which an outcome becomes registered (the detection or measurement) is defined with relation to the capabilities of the human observer (the way the human brain functions after all). Physicians define death as the ``irreversible" break-down of all the brain functions included brainstem. What do they mean by ``irreversible"? Just that a damage happens beyond our capabilities to repair. Similarly I think that at detection something happens beyond our capabilities to restore.

Actually any science based on observations has necessarily to define the observer who observes. For the science we know and do, observation relates to the human observer. In particular, it is often overlooked that the observer problem affects relativity as well:

In special relativity, one invokes the ``grand-father" paradox to exclude signals traveling faster than light, that is, one argues that such signals could be used to kill one's grand-father and this would be absurd. But why is this absurd? Actually because we assume that an observation is something irreversible and cannot be considered unhappened: Evidence about one's grand-father existence cannot be reversed to non-evidence. But special relativity doesn't tell us the reason for this irreversibility, and this means that also this theory has to struggle with the ``measurement problem".

By reducing gravity to a geometric property of space-time, general relativity unavoidably involves singularities and ``carries within itself the seeds of its own destruction." The singularity problem is related to the observer problem: At any black hole there is a ``world's end" beyond which it does not make sense to speak of observation; the ``world's end" is a hyper-surface defined by either the ``event-horizon" or the ``death-horizon", depending on the size of the black hole.\cite{arsu} Accordingly there is no ``real physical space-time" (there are no ``correlated \emph{visible} things") in the region within the ``world's end", and then there is no ``real physical" black hole singularity either.

The "measurement problem" is in a sense the physical correlate of the ``halting problem" in arithmetics. However in arithmetics we can sharply prove that at any time T there will be questions about numbers that we cannot answer with the methods available at time T (the well known Turing's theorem). By contrast in physics we cannot yet sharply prove (we only intuitively feel) that at any time T there will be physical processes (for instance death) that lie beyond our capabilities to restore, and therefore are irreversible \emph{with relation to the human capabilities}. For the time being the conditions defining this irreversibility (likely involving a new constant of nature) are unknown: this is the ``measurement problem" and shows that to date both quantum physics and relativity are incomplete.

\ \\
\textbf{Conclusion}.\textemdash Both, quantum nonlocal correlations and relativistic local ones, assume ``free will" on the part of the experimenter, and mean that observed events are correlated without ``particles" traveling through trajectories in space-time, that is, without continuous connection in between. This may be the very meaning of the ``space-time quantization". The introduction of the space-time continuum (geometry and ``real" numbers) in physics is a useful idealization, but it should not be considered a ``real" structure underpinning the physical world.\cite{ng}

\ \\
\emph{Acknowledgments}: I acknowledge discussions with the participants to the ``Workshop on Time in Physics", ETH Zurich, September 7-11, 2015, in particular {\v C}aslav Brukner,  J{\"u}rg Fr{\"o}hlich, Thomas Pashby, Sandu Popescu, Renato Renner, Vlatko Vedral, and
Stefan Wolf.

\end{document}